\documentclass[submission,copyright,creativecommons,sharealike]{eptcs}
\providecommand{\event}{MARS 2017} % Name of the event you are submitting to
\usepackage{underscore}           % Only needed if you use pdflatex.
\usepackage{graphicx}
\graphicspath{ {images/} }
\usepackage{array}   % for tables

\title{A Benchmark on Reliability of Complex\\
 Discrete Systems: Emergency Power Supply\\
 of a Nuclear Power Plant
}
\author{Marc Bouissou
\institute{EDF Lab Saclay\\ 7 bd Gaspard Monge,\\ 
Palaiseau 91120, France}
\email{Marc.Bouissou@edf.fr}
}
\def\titlerunning{A Benchmark on Reliability of Complex Discrete Systems}
\def\authorrunning{Marc Bouissou}
\begin{document}
\maketitle

\begin{abstract}
Abstract: This paper contains two parts: the description of a real electrical system, with many redundancies, reconfigurations and repairs, then the description of a reliability model of this system, based on the BDMP (Boolean logic Driven Markov Processes) formalism and partial results of a reliability and availability calculation made from this model. 
\end{abstract}

\section{Introduction}

EDF is launching in 2017 a benchmark on methods and tools for the dependability assessment of dynamic stochastic systems. This benchmark will include several test cases, with graded difficulty. Some test cases will be purely discrete systems, some others will include hybrid aspects. The test case described in this paper is a discrete system, including most difficulties one can encounter when assessing the dependability of a complex system: reconfigurations with cascades of instantaneous probabilistic transitions, repairs, high redundancy level, common cause failures, large differences between the lowest and highest transition rates, multidirectional interactions (because of short circuits), looped interactions, existence of deterministic delays (due to battery depletion). For a system of this kind, by dependability we mean both \emph{reliability} (the probability R(t)  that the system performs a required function without interruption during the time interval [0, t]) and \emph{availability} (the probability A(t) that the system performs a required function at a given time). The system described in the next section is taken from a French nuclear power plant. The hypotheses on the qualitative behaviour of the components and the control system are as accurate as possible, but the reliability data (failure and repair rates) are fake, for confidentiality reasons. However the orders of magnitude are realistic.

\section{System description and benchmark goal}

This emergency power supply is a critical system whose goal is to provide energy to components essential to the safety of a nuclear power plant, mainly the instrumentation and control systems, the pumps of the primary circuit (required to function at all times, whatever the state of the plant) and the pumps of various cooling circuits, used both in normal and incidental/accidental situations. It is an electric power system with repairable components, failures in function and on demand, cold redundancies, and common cause failures. Other descriptions (much simpler than this one) of the system to be studied can be found in \cite{BouissouBon2003} and \cite{Brameretthesis2015}. 

\subsection{System functioning and undesirable event}
\begin{figure}[h]
\includegraphics[width=0.9\textwidth]{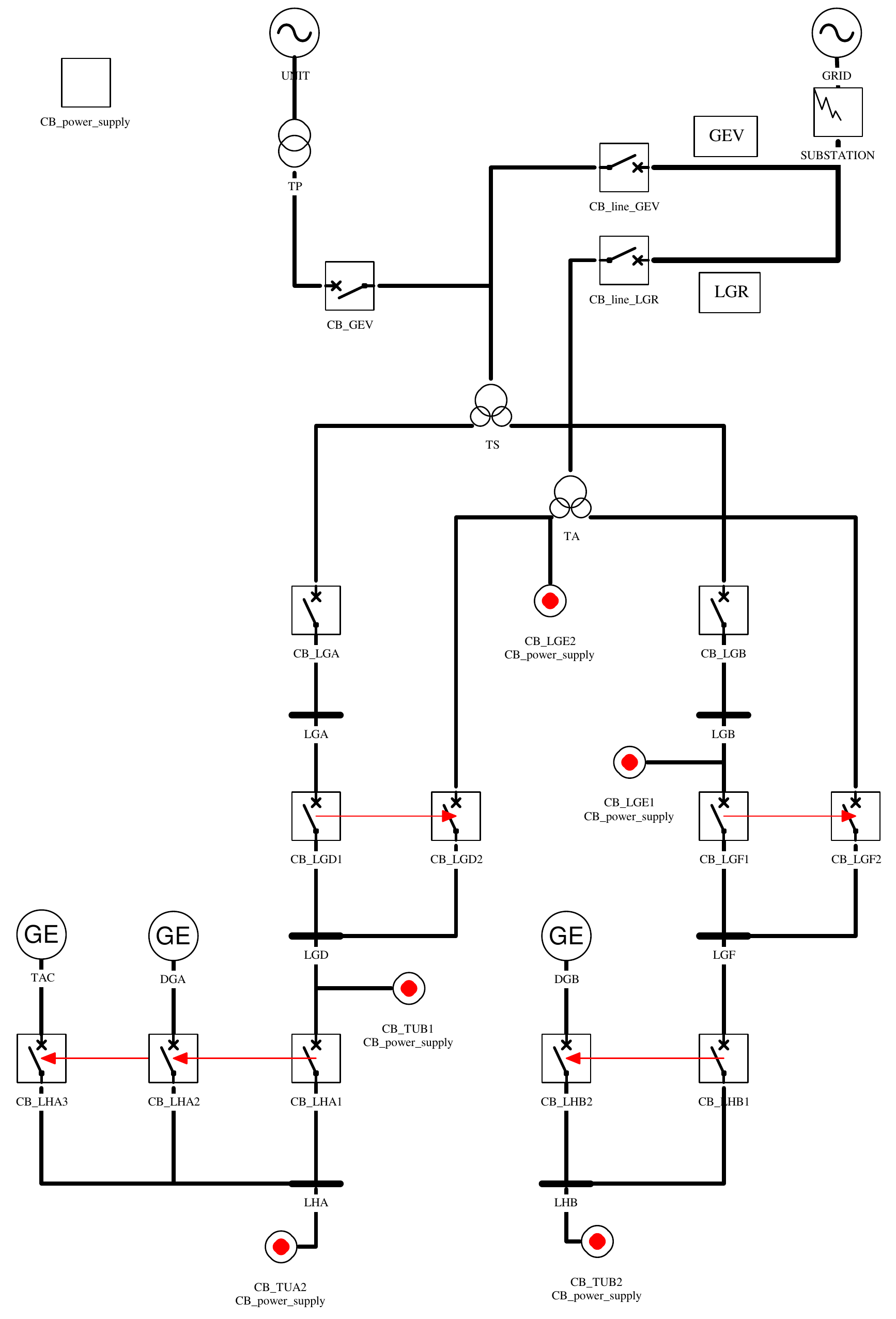}
\centering
\caption{High voltage part. The red arrows represent the reconfiguration strategy}
\label{fig:hv}
\end{figure}

\begin{figure}[h]
\includegraphics[width=0.9\textwidth]{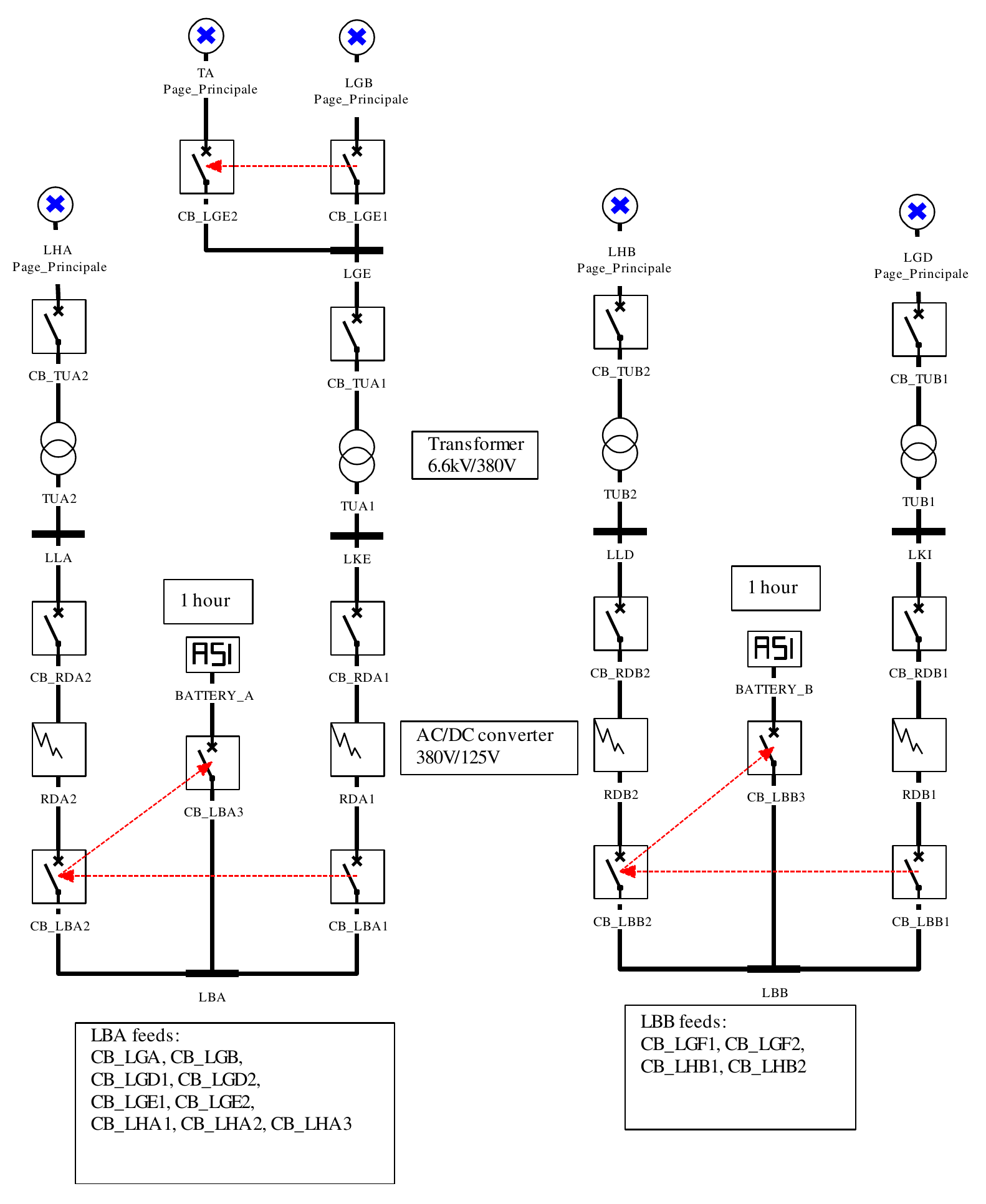}
\centering
\caption{Low voltage part. The red arrows represent the reconfiguration strategy}
\label{fig:lv}
\end{figure}

The physical description of the system is given in Fig.~\ref{fig:hv} and Fig.~\ref{fig:lv}. The role of the system is to supply electricity out of the bus bars LHA and LHB (cf.\ Fig.~\ref{fig:hv}). The regular power supply of bus bars LHA and LHB comes from the transformer TS. TS is supplied by the GRID or by the plant UNIT. When the GRID is available, UNIT works in regular mode and injects power into the GRID. If UNIT fails, the GRID can still feed the system through TS. 
When the GRID fails, or an element on the path from CB\_line\_GEV to GRID, there is an attempt to switch to house load mode, where the power of UNIT is reduced, to feed only the plant itself. This mode is rather unstable. The probability of a successful reconfiguration to house load mode is relatively low, and the failure rate\footnote{The failure rate of an item (component or system) is defined formally as: - R'(t)/R(t). When the time to failure of this item is exponentially distributed with a mean time to failure equal to MTTF, its reliability R(t) = exp(- t/MTTF) and its failure rate is constant and equal to 1/MTTF.}  in function of this mode is high (cf.\ Section \ref{reldata}). After a failure of the house load mode, it is impossible to restart UNIT (and thus restore the house load mode) until the path from UNIT to GRID is completely repaired and the GRID is available. LHA and LHB can also be powered by the GRID through transformer TA. Diesel generators DGA and DGB supply respectively LHA and LHB when these bus bars are not powered by LGD and LGF. The Diesel generators may refuse to start and have two kinds of failure in function, called ``short'' and ``long'' because they correspond to very different repair rates.
A last resort power source called TAC 
(for ``turbine \`a combustion'') can be used to power LHA when all other sources (including DGA) are lost.  
All components except sources may have short circuits. Circuit breakers (CB\_*) may refuse to close or to open and may have short circuits. The circuit breakers need an auxiliary supply to be able to open and close; this supply (125V) is provided by the bus bars LBA and LBB. 
NB: for the sake of simplicity, all circuit breakers except those listed at the bottom of Fig.~\ref{fig:lv} will be considered as ``perfectly powered''€.
There are common cause failures on lines (GEV and LGR) and on diesel generators. Since the two lines GEV and LGR are connected to the same part of the substation, a short circuit on one of them also makes the other one unavailable. This failure mode of both lines can be repaired quickly, but they may also be lost because of extreme weather conditions, and in this case the repair time is long.

The problem to be solved in this benchmark is to compute the unreliability and unavailability\footnote{The reliability and availability of this system are so close to 1, that it is more convenient to work with their complements to 1, called respectively unreliability and unavailability.}   of the system, for a mission time of 10000 hours, i.e. about one year. The undesirable event to be quantified is the loss of power on both LHA and LHB, because owing to redundancies the loss of only one bus bar does not affect the important functions of the power plant.  In addition, the important contributors to unreliability (resp. unavailability) must be identified, with the objective of building a simplified model able to account for 80\% of the unreliability and 80\% of the unavailability.

\subsection{Initial state}
The initial state is the ``€œperfect state''€ where all components are working. It is the state the system always returns to thanks to repairs. In this state, UNIT produces power which is sent towards GRID, and from which a part is withdrawn in order to feed LHA and LHB. All circuit breakers are closed, except the following: 
\begin{itemize}
\item High voltage: CB\_LHA3, CB\_LHA2, CB\_LGD2, CB\_LGF2, CB\_LHB2,
\item Low voltage: CB\_LGE2, CB\_LBA2, CB\_LBA3, CB\_LBB2, CB\_LBB3.
\end{itemize}

\subsection{Reliability data} \label{reldata}
The table below gives the list of failure modes to be taken into account for the various components of the system, with the corresponding numerical data.

\begin{center}
\begin{tabular}{ || m{7cm} | m{2cm} | m{2cm} | m{2cm} || } 
\hline
Component type/failure mode & gamma Probability of failure on demand & lambda/h Failure rate (constant) & mu/h Repair rate (constant) \\ 
\hline Circuit breaker/refuse to open (all voltages) & 2.00E-04 &  & 1/5 \\
\hline Circuit breaker/refuse to close (all voltages) & 2.00E-04 &  & 1/5 \\
\hline CB_GEV, CB_line_GEV, CB_line_LGR short circuit &  & 1.00E-07 & 1/5 \\
\hline Circuit breaker/short circuit (all other high voltage circuit breakers) &  & 5.00E-07 & 1/5 \\
\hline Circuit breaker/short circuit (all low voltage circuit breakers) &  & 1.00E-06 & 1/5 \\
\hline Bus bar short circuit (high voltage) &  & 2.00E-07 & 1/50 \\
\hline Bus bar short circuit (low voltage) &  & 5.00E-07 & 1/50 \\
\hline Transformer short circuit (TP, TS, TA) &  & 5.00E-06 & 1/200 \\
\hline Transformer short circuit (TUA1, TUA2, TUB1, TUB2) &  & 2.00E-07 & 1/10 \\
\hline Diesel generators/long failure & 2.00E-03 & 5.00E-04 & 1/200 \\
\hline Diesel generators/short failure &  & 2.00E-03 & 1/10 \\
\hline TAC & 2.00E-03 & 1.00E-03 & 1/200 \\
\hline GRID failure in function &  & 1.00E-05 & 1/10 \\
\hline UNIT (normal operation) failure in function &  & 1.00E-04 & 1/10 \\
\hline UNIT (house load operation) & 0.2 & 0.1 & 1/20 after GRID repair \\
\hline SUBSTATION &  & 1.00E-06 & 1/20 \\
\hline Lines GEV, LGR/short circuit &  & 2.00E-05 & 1/5 \\
\hline AC/DC converter (RDA1, RDA2, RDB1, RDB2) &  & 1.00E-06 & 1/3 \\
\hline Simultaneous failure of DGA and DGB by CCF & 2.00E-04 & 5.00E-05 & 1/400 \\
\hline Simultaneous failure of GEV and LGR by CCF due to bad climatic conditions &  & 1.00E-06 & 1/200 \\
\hline
\end{tabular}
\end{center}
\paragraph{Common cause failures (CCF)}
In order to be pessimistic, the mean time needed to repair completely the two diesel generators after a common cause failure is considered to be twice the time needed for one. It is also possible to represent two independent repairs, each one with a mean duration of 200 hours. 
For the lines, the situation is different: the most likely reason for a CCF on the two lines is extreme climatic conditions, which means that the repair is likely to be much longer than the repair of two ``€œordinary'' failures of the lines.

\subsection{More details on failure modes and reconfigurations}
\subsubsection {Short circuit ``€œpropagation''€}
A short circuit is the (unwanted) connection between an electrical component and the ground. This induces a very high current intensity between the electrical sources and this component. Normally, circuit breakers immediately open in order to disconnect the short circuit from sources and to avoid excessive heat and components destructions. Short circuits propagate towards sources in the sense that until sources are disconnected, all components between sources and the short circuit are traversed by a high intensity. In fact their effects propagate in all directions. Consider, for example, the bus bars LGA and LGD. If a short circuit happens on LGA, and CB\_LGA refuses to open, the transformer TS can be damaged. Furthermore, another consequence must be taken into account: if CB\_LGD1 does not open, it becomes impossible to feed LGD by closing CB\_LGD2 because this would establish a new path between a source and the short circuit.

\subsubsection{Circuit breakers control}
An essential point in the dependability of the system is the multiple reconfigurations that can be done automatically thanks to circuit breakers and their control. The structure of the system and its control are designed in such a way that all failures (except short circuits on LHA and LHB themselves) can trigger the opening of circuit breakers in order to isolate the failed component and closure of other circuit breakers in order to feed again LHA and LHB. 
The structure of the system makes it possible to specify all these reconfigurations by a local description of priorities, at each bus bar that can be fed by different sources. These priorities are represented by red dotted lines (the ``triggers''). For example, for feeding LHA, the priority order is: by CB\_LHA1, then if this is impossible, by CB\_LH2, then if the two previous ones are impossible by CB\_LHA3.
Of course, there are physical components behind this, but since they are very reliable and the effect of their failures is equivalent to failure to open or failure to close of circuit breakers, they are not specified precisely in the definition of the test case. 
If this is convenient for certain models, one can consider that each set of ``€œtriggers''€ corresponds to an automaton that monitors the presence of tension on each branch and the working/failure status of circuit breakers, and sends to each circuit breaker the orders corresponding to the priorities specified by triggers. In order not to start diesel generators when it is not needed, these automatons effectively send orders to components after a short delay, and the delays are chosen such that the reconfigurations take place first at the PLANT/GRID level then TS/TA transformers, then diesel generators.

\section{Modeling the system with BDMP}
\subsection{BDMP in a nutshell}
The general idea of BDMP \cite{BouissouBon2003}, as suggested by their name, is to associate a Markov process (which represents the behavior of a component or a subsystem) to each leaf of a fault-tree. This fault-tree is the structure function of the system. 
What is really new with BDMP is that:
\begin{itemize}
\item the basic Markov processes have two ``modes'', corresponding to the fact that the components or subsystems that they model are required or are in standby (of course, they can also have only one mode, and the meaning of the modes may be different in some cases),
\item at any time, the choice of the mode of one of the Markov processes (unless it is independent) depends on the value of a Boolean function of other processes.
\end{itemize}

\begin{figure}[h]
\includegraphics[width=0.3\textwidth]{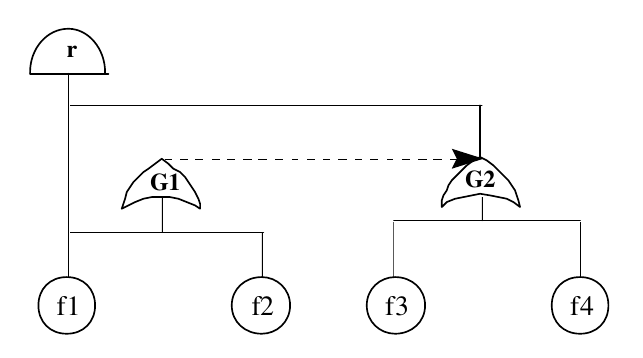}
\centering
\caption{a simple BDMP}
\label{fig:smallbdmp}
\end{figure}

An extreme case is when the processes are independent. This corresponds to a fault-tree, the leaves of which are associated to independent Markov processes.
The dependencies between parts of a BDMP are specified by ``€œtriggers''€. 
A trigger is represented graphically with a dotted line. Two triggers must not have the same target. This means that it is sometimes necessary to create an additional gate (like G1 in Fig.~\ref{fig:smallbdmp}) whose only function is to define the origin of a trigger. Figure \ref{fig:smallbdmp} is an example of graphical representation of all the notions of BDMP. In this example, we have a fault-tree with two tops: r (the main one) and G1. The basic events are f1, f2, f3, and f4: they represent standard triggered Markov processes: failures than can happen only in the ``€œrequired''€ mode. There is only one trigger, from G1 to G2. This trigger makes the mode of f3 and f4 switch from ``€œnon required''€ to ``€œrequired''€ whenever G1 is true.
For a complete mathematical definition of BDMP, refer to \cite{BouissouBon2003}.

\subsection{Tools used}
In order to build the BDMP model and to make calculations from it we have used the EDF tools of the Figaro workbench. These tools have been constantly improved since their first versions delivered in 1990 \cite{Bouissou1991}. 
The Figaro language, created in 1990, is a (free and public) domain specific object oriented modeling language dedicated to dependability. It generalizes all the usual reliability models, and can easily be associated to various graphical representations. It allows to cast generic models in knowledge bases (KB). A formal definition of its semantics is available in \cite{consistency}. The Figaro workbench comprises a set of tools to create Figaro models and to process them in order to perform dependability analyses. Here is a list of the main tools:
\begin{itemize}
\item FigaroIDE is an integrated development environment for creating KBs;
\item KB3 is a generic graphical user interface. Once a KB has been loaded in KB3, it becomes a specialized GUI for building a certain kind of graphical models. KB3 comes with a few ``€œabstract KBs''€ corresponding to classical reliability models: reliability block diagrams, digraphs, Petri nets, BDMP€\dots. KB3 provides sophisticated functions to input and manage complex system models, perform interactive simulations and can generate fault trees and display them graphically;
\item Figseq is a quantification tool that explores the sequences leading to a target state, defined by a Boolean expression. Given the mission time and truncation criteria, Figseq computes an estimated value and an upper bound of the undesirable event probability. It can perform reliability and availability calculations;
\item YAMS is another solver: it uses Monte Carlo simulation on the system model to compute various quantities, including reliability and availability. Any kind of probability distribution can be associated to transitions with this tool. YAMS is also able to output a selection of simulated scenarios, but the obtained results are much more ``€œnoisy''€ than those obtained with Figseq.
\end{itemize}
 All the tools cited above are general, and can be used for any kind of system. In order to specialize them and create a powerful tool dedicated to the study of electrical systems, it is only needed to write a Figaro knowledge base. And indeed, such a KB (called K6) was created at EDF in order to automate the study of electrical networks like those of factories, Internet data centers, hospitals etc. \cite{K6}. With K6, it is possible to input graphically the physical layout of an electrical system in KB3, then simulate interactively its behavior (with color changes on the system picture, according to state changes) in order to validate the model. Then dependability measures can be obtained by Figseq and YAMS. However we could not have used K6 without making significant changes in this KB to solve the test case described above. This is why we used K6 only in order to draw the system layout (Figures \ref{fig:hv} and \ref{fig:lv}) but  preferred to use the BDMP KB, which is domain independent and did not require any change, to do the study. 
Because of the lack of space we cannot give much information on the tools here, but the paper \cite{CIEM} gives a good overview of the main tools of the Figaro workbench. 

We emphasize that all the figures in this paper except Fig.~\ref{fig:smallbdmp} were directly copied and pasted from the KB3 graphical user interface.    

\subsection{The system model}
The BDMP graphical representation available in appendix contains all what is necessary to understand the model contents. The only information that is not visible in the pictures is the reliability data. But this data simply reflects the contents of Section \ref{reldata}. As a complement to this ``high level'' description of the BDMP, we provide the text file containing the Figaro description (generated from the graphics by KB3) that was the input of the Figseq solver. 

We made a few simplifying assumptions in order to keep the model relatively simple.
We supposed that the sources of the low voltage part (which, in fact come from the high voltage part except the battery) are perfectly reliable.
We replaced the fixed time (one hour) of the battery depletion by an Erlang distribution: Erl(2, 2/h). This is necessary to keep a markovian model and to be able to use the tool Figseq which uses analytical formulae. It is easy to put a fixed time transition in the model, but then the only available quantification method becomes Monte Carlo simulation, with much less interesting qualitative results and computation times depending heavily on the order of magnitude of the probabilities to be estimated.   
In order to model the fact that the house load functioning cannot be repaired until the GRID is available, we created a single repairman who repairs the basic events relative to the GRID and the house load functioning. Since from the initial state, the house load functioning is not active, this repairman will always be taken first to repair the GRID. This will inhibit the repair of the house load functioning until the GRID is repaired.

\section{A few results}
Since we would like other researchers to try to solve this benchmark, we will give here only a small part of the results we are able to obtain with Figseq, the Markov analysis tool that we use for the quantification of KB3 models in general, and BDMPs in particular. 
Figseq explores paths in the Markov graph implicitly defined by the BDMP. This graph is huge and it is impossible to build it exhaustively, even on a very powerful machine. But most systems, even very reliable ones, have some weak points. The exploration of all sequences having a probability larger than a given threshold can reveal those weak points and save the work of exploration of sequences negligible compared to dominant sequences.
Figseq is based on a smart algorithm that explores loops in the Markov graph only once, while taking into account in the probability calculation the fact that these loops can be traversed any number of times. 
But what makes the exploration so efficient (a result with a precision of about 10\% on the reliability and asymptotic availability obtained on a simple laptop in 5mn) is the properties of BDMP that reduce drastically the number of sequences to explore thanks to the ``€œrelevant event filtering''€ technique, explained in \cite{BouissouBon2003}. 

The most probable sequence found with this model is the following:
\begin{center}
\begin{tabular}{ || m{5cm} | m{2cm} | m{2cm} || } 
\hline	failF(CCF_GEV_LGR) &	1e-6 &	EXP \\
\hline	good(on_demand_house) &	0.8	& INS \\
\hline	failF(in_function_house)& 	0.1 &	EXP  \\
\hline	good(demand_CCF_DG)	& 0.9998 &	INS  \\
\hline	good(demand_DGA) & 0.998 & INS  \\
    	good(demand_DGB) & 0.998 & INS	 \\
\hline	good(RO_CB_LHA1) & 	0.9999 &	INS  \\
    	good(RO_CB_LHB1)& 	0.9999 &	INS  \\
\hline	good(RC_CB_LHA2)  & 	0.9999 &	INS  \\
    	good(RC_CB_LHB2) & 	0.9999 &	INS  \\
\hline	failF(CCF_DG) & 	5e-5 &	EXP  \\
\hline	good(demand_TAC) &	0.998 & 	INS  \\
\hline	good(RO_CB_LHA2) &	0.9998 &	INS  \\
\hline	good(RC_CB_LHA3) &	0.9995 &	INS \\
\hline	failF(TAC) & 	0.001 &	EXP  \\
\hline
\end{tabular}
\end{center}

In this table, the two last columns give the characteristics of transitions: exponential (with rate in $h^{-1}$) or instantaneous. In this sequence, the initiator is the common cause failure on the lines (this has a long repair time, which explains why all the other events of the sequence can take place). The switch to house load is successful but then the house load fails in operation. So the diesel generators start, the circuit breakers allowing to use them as sources open and close normally. Then the two diesel generators are lost because of a common  cause failure in function; the last possible reconfiguration (to use the TAC) succeeds and finally the TAC fails in function.

NB: the absence of line between, for example, good(demand\_DGA) and good(demand\_DGB) shows that Figseq considered those two instantaneous events as simultaneous and independent (they concern different state variables of the model).

Hereafter is another example of sequence, with a very small contribution ($< 0.0001$). In this case, no common cause failure is involved, but we have independent failures of, successively the two transformers TS and TA, then of diesel generators and TAC.
The possibility to obtain sequences is a major advantage over other Markov analysis techniques: it is indeed the only way to gain confidence in the model.

\begin{center}
\begin{tabular}{ || m{5cm} | m{2cm} | m{2cm} || } 
\hline failF(TS) &	5e-6 & EXP \\
\hline good(RC_CB_LGD2) & 0.9995 & INS \\
       good(RC_CB_LGF2)  & 0.9995 & INS \\

\hline failF(TA) &	5e-6 & EXP \\
\hline good(demand_CCF_DG) & 	0.9998 & INS \\
\hline good(demand_DGA) &	0.998 & INS \\
       good(demand_DGB) &	0.998 & INS \\

\hline good(RO_CB_LHA1) &	0.9999  & INS \\
       good(RO_CB_LHB1) &	0.9999  & INS \\

\hline good(RC_CB_LHA2) &	0.9999  & INS \\
       good(RC_CB_LHB2) &	0.9999  & INS \\

\hline failF(DGA_long)&	0.0005  & EXP \\
\hline good(demand_TAC)	& 0.998 & INS \\
\hline good(RO_CB_LHA2)	& 0.9998 & INS \\
\hline good(RC_CB_LHA3)	& 0.9995 & INS \\
\hline failF(DGB_short)	& 0.002  & EXP \\
\hline failF(TAC)	& 0.001  & EXP \\

\hline
\end{tabular}
\end{center}

\vspace*{-1mm}
\section{Conclusion}
\vspace*{-1mm}
In this paper we have described a real problem of dependability assessment, as those encountered in the industry. The relative imprecision of the description in English that is given in addition to the topology of the system is the everyday life of reliability analysts. They also usually have a hard time looking for reliability data. But in order to make results of the benchmark comparable, we have given a precise list of failure modes and the associated numerical data as part of the problem definition. 
We have also given a first formal model of this system, based on the BDMP mathematical formalism. 
Finally we have given small excerpts of results obtained with this model, without unveiling the global results. It will be interesting to look at those results when solutions obtained with some other tools are available. Any person interested in solving this benchmark is strongly encouraged to contact the author of this paper!

\nocite{*}
\vspace*{-2mm}
\bibliographystyle{eptcs}
\bibliography{generic}

\begin{thebibliography}{1}
\providecommand{\bibitemdeclare}[2]{}
\providecommand{\surnamestart}{}
\providecommand{\surnameend}{}
\providecommand{\urlprefix}{Available at }
\providecommand{\url}[1]{\texttt{#1}}
\providecommand{\href}[2]{\texttt{#2}}
\providecommand{\urlalt}[2]{\href{#1}{#2}}
\providecommand{\doi}[1]{doi:\urlalt{http://dx.doi.org/#1}{#1}}
\providecommand{\bibinfo}[2]{#2}

\bibitemdeclare{misc}{bibliographystylewebpage}
\bibitem{bibliographystylewebpage}
\emph{\bibinfo{title}{EDF dependability analysis tools}}.
\newblock \urlprefix\url{http://sourceforge.net/projects/visualfigaro/}.

\bibitemdeclare{inproceedings}{CIEM}
\bibitem{CIEM}
\bibinfo{author}{M.~\surnamestart Bouissou\surnameend} (\bibinfo{year}{2005}):
  \emph{\bibinfo{title}{Automated dependability analysis of complex systems
  with the KB3 workbench: the experience of EDF R\&D}}.
\newblock In: {\sl \bibinfo{booktitle}{CIEM 2005}},
  \bibinfo{address}{Bucharest}.

\bibitemdeclare{article}{BouissouBon2003}
\bibitem{BouissouBon2003}
\bibinfo{author}{M.~\surnamestart Bouissou\surnameend} \&
  \bibinfo{author}{J.-L. \surnamestart Bon\surnameend} (\bibinfo{year}{2003}):
  \emph{\bibinfo{title}{A new formalism that combines advantages of fault-trees
  and Markov models: Boolean logic driven Markov processes}}.
\newblock {\sl \bibinfo{journal}{Reliability Engineering and System Safety}}
  \bibinfo{volume}{Vol. 82}, pp. \bibinfo{pages}{149--163},
  \doi{10.1016/S0951-8320(03)00143-1}.

\bibitemdeclare{inproceedings}{Bouissou1991}
\bibitem{Bouissou1991}
\bibinfo{author}{M.~\surnamestart Bouissou\surnameend},
  \bibinfo{author}{H.~\surnamestart Bouhadana\surnameend},
  \bibinfo{author}{M.~\surnamestart Bannelier\surnameend} \&
  \bibinfo{author}{N.~\surnamestart Villatte\surnameend}
  (\bibinfo{year}{1991}): \emph{\bibinfo{title}{Knowledge modelling and
  reliability processing: presentation of the FIGARO language and associated
  tools}}.
\newblock In: {\sl \bibinfo{booktitle}{Safecomp'91}},
  \bibinfo{address}{Trondheim (Norway)}.

\bibitemdeclare{inproceedings}{consistency}
\bibitem{consistency}
\bibinfo{author}{M.~\surnamestart Bouissou\surnameend} \&
  \bibinfo{author}{J.-C. \surnamestart Houdebine\surnameend}
  (\bibinfo{year}{2002}): \emph{\bibinfo{title}{Inconsistency detection in KB3
  models}}.
\newblock In: {\sl \bibinfo{booktitle}{ESREL 2002}}, \bibinfo{address}{Lyon
  (France)}.

\bibitemdeclare{phdthesis}{Brameretthesis2015}
\bibitem{Brameretthesis2015}
\bibinfo{author}{P.-A. \surnamestart Brameret\surnameend}
  (\bibinfo{year}{2015}): \emph{\bibinfo{title}{Assessment of reliability
  indicators from automatically generated partial Markov chains}}.
\newblock Ph.D. thesis, \bibinfo{school}{LURPA}, \bibinfo{address}{{\'E}cole
  normale sup{\'e}rieure de Cachan}.

\bibitemdeclare{inproceedings}{K6}
\bibitem{K6}
\bibinfo{author}{T.~\surnamestart Chaudonneret\surnameend},
  \bibinfo{author}{T.~\surnamestart Moreau\surnameend} \&
  \bibinfo{author}{C.~\surnamestart Monnier\surnameend} (\bibinfo{year}{2016}):
  \emph{\bibinfo{title}{L'outil K6 pour les {\'e}tudes de s{\^u}ret{\'e} de
  fonctionnement des r{\'e}seaux {\'e}lectriques industriels}}.
\newblock \bibinfo{publisher}{Colloque Lambda-mu 2017}, \bibinfo{address}{St
  Malo (France)},
  \doi{10.4267/2042/61804}.

\end{thebibliography}

\clearpage % forces float figures to be printed before this point
\appendix
\section{Appendix}
The graphical BDMP model is divided in 6 pages in the KB3 tool. They are reproduced in the following pages. The full description of the BDMP written in the Figaro 0 language (sub language of Figaro used as input for solvers Figseq and YAMS) is available as a separate ASCII file.
Here are some explanations about the meaning of the various kinds of links and symbols used in the graphical representation. 

\begin{itemize}
\item Red arrows represent triggers.

\item Grey arrows represent constraints in the order of instantaneous events. For example, they force the test on startup of diesel generators to be done before (instead of simultaneously) the circuit breakers reconfigurations.

\item The OR gates with the symbol of approximation are used to aggregate their input leaves into a single leaf that has the sum of failure rates of the input leaves as failure rate, and a weighed sum of the repair rates of the input leaves as repair rate. This reduces very much the combinatorial explosion, at the cost of an approximation that is acceptable provided the repair rates are not too disparate.

\item The AND gates with a triangle change from False to True when their second input changes from False to True while their first input is True. They are used in the Circuit\_breaker pages to model the fact that if there is a need to change the position of circuit breakers while their low voltage power supply is lost, they will not respond.

\item Small circles with a red point or a blue cross represent split links. They have their origin (the blue cross) in a page and target (the red point) in another page. Sometimes split links are also used within a single page for aesthetic reasons.

\end{itemize}

% Attention! there must be no "_" in the file names

\begin{figure}[h]
\includegraphics[width=0.9\textwidth]{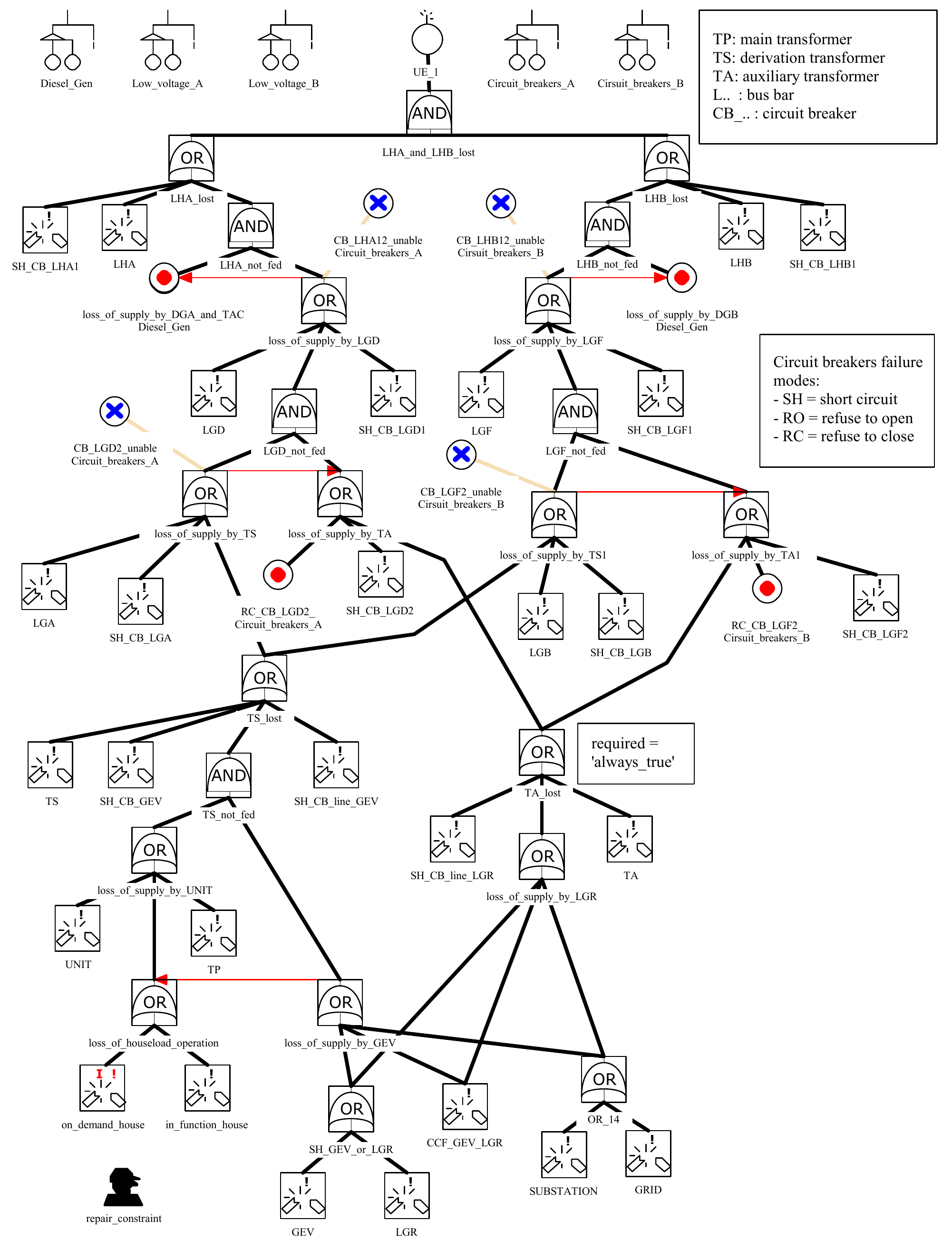}
\centering
\caption{Page Main\_page. }
\end{figure}

\begin{figure}[h]
\includegraphics[width=0.9\textwidth]{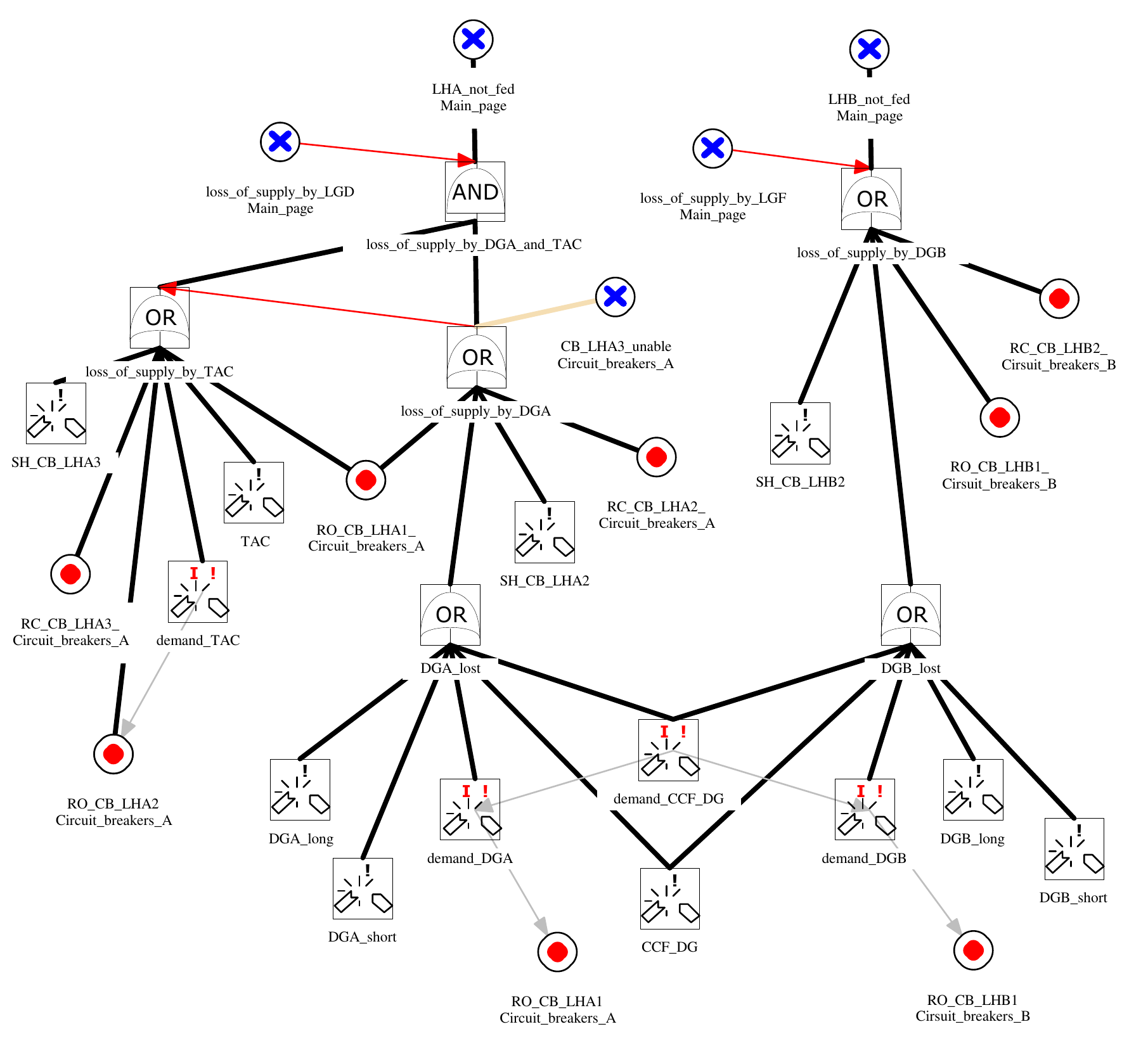}
\centering
\caption{Page Diesel\_Gen. }
\end{figure}

\begin{figure}[h]
\includegraphics[width=0.9\textwidth]{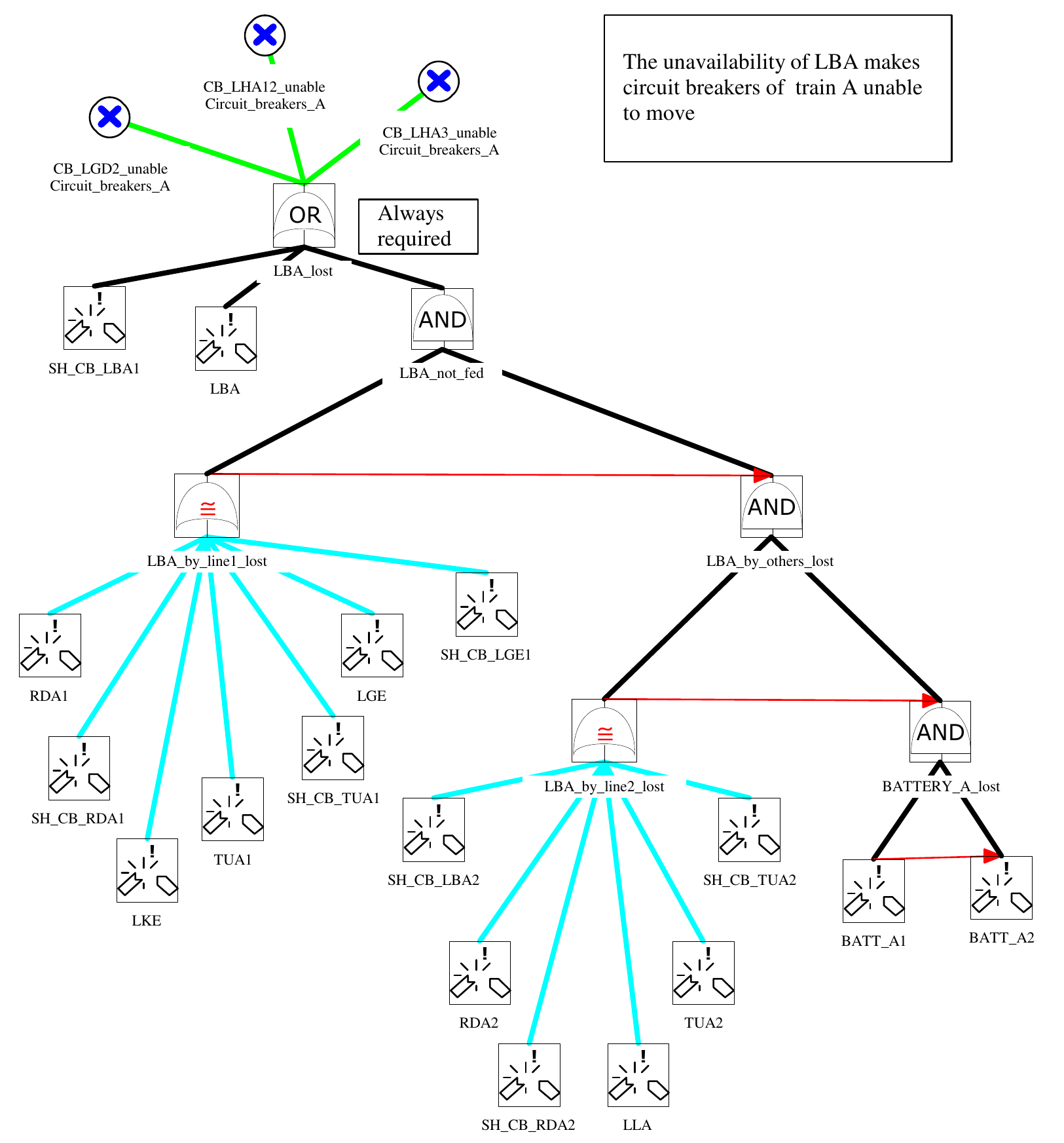}
\centering
\caption{Page Low\_voltage\_A.}
\end{figure}

\begin{figure}[h]
\includegraphics[width=0.9\textwidth]{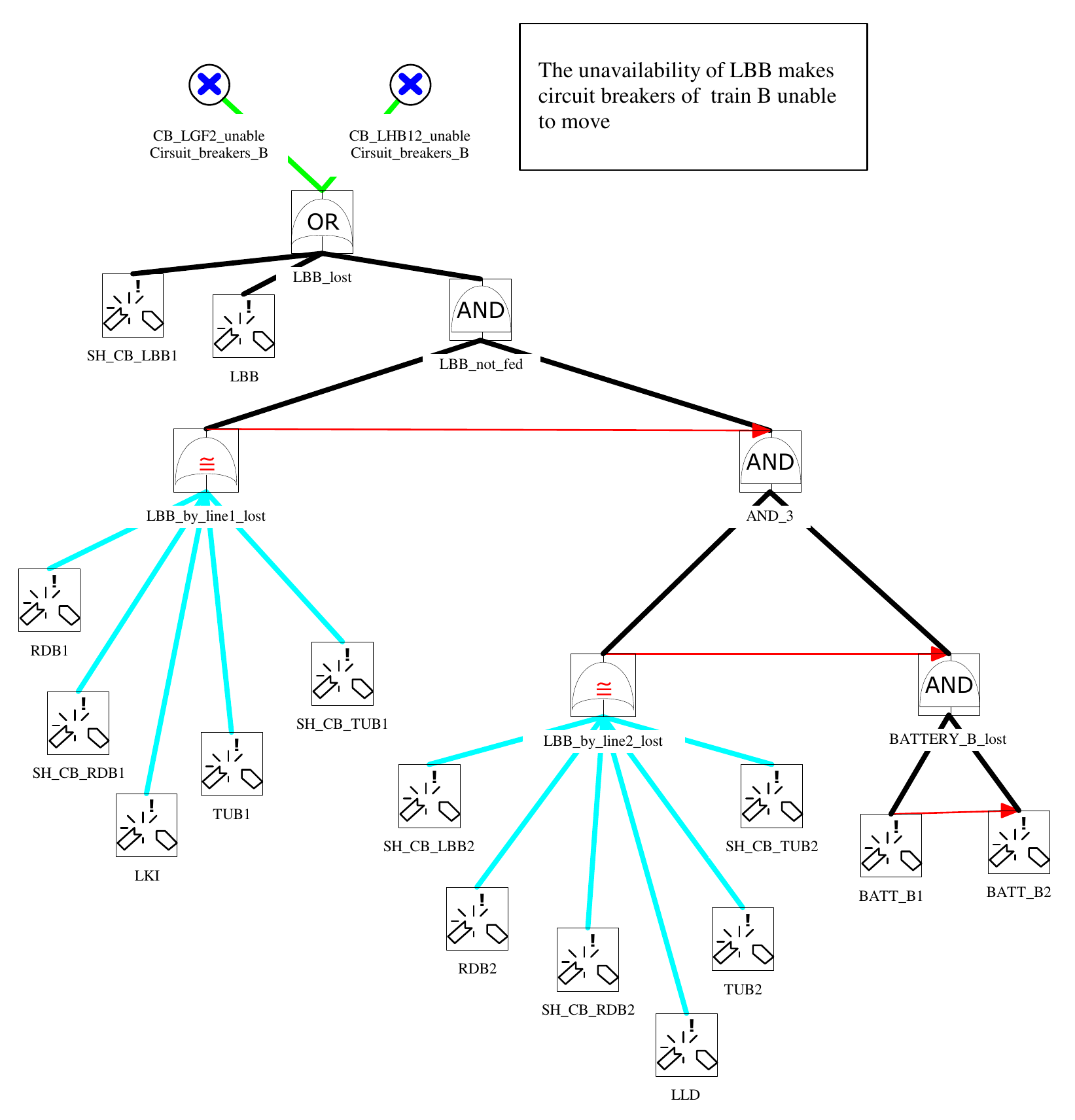}
\centering
\caption{Page Low\_voltage\_B.}
\end{figure}

\begin{figure}[h]
\includegraphics[width=0.9\textwidth]{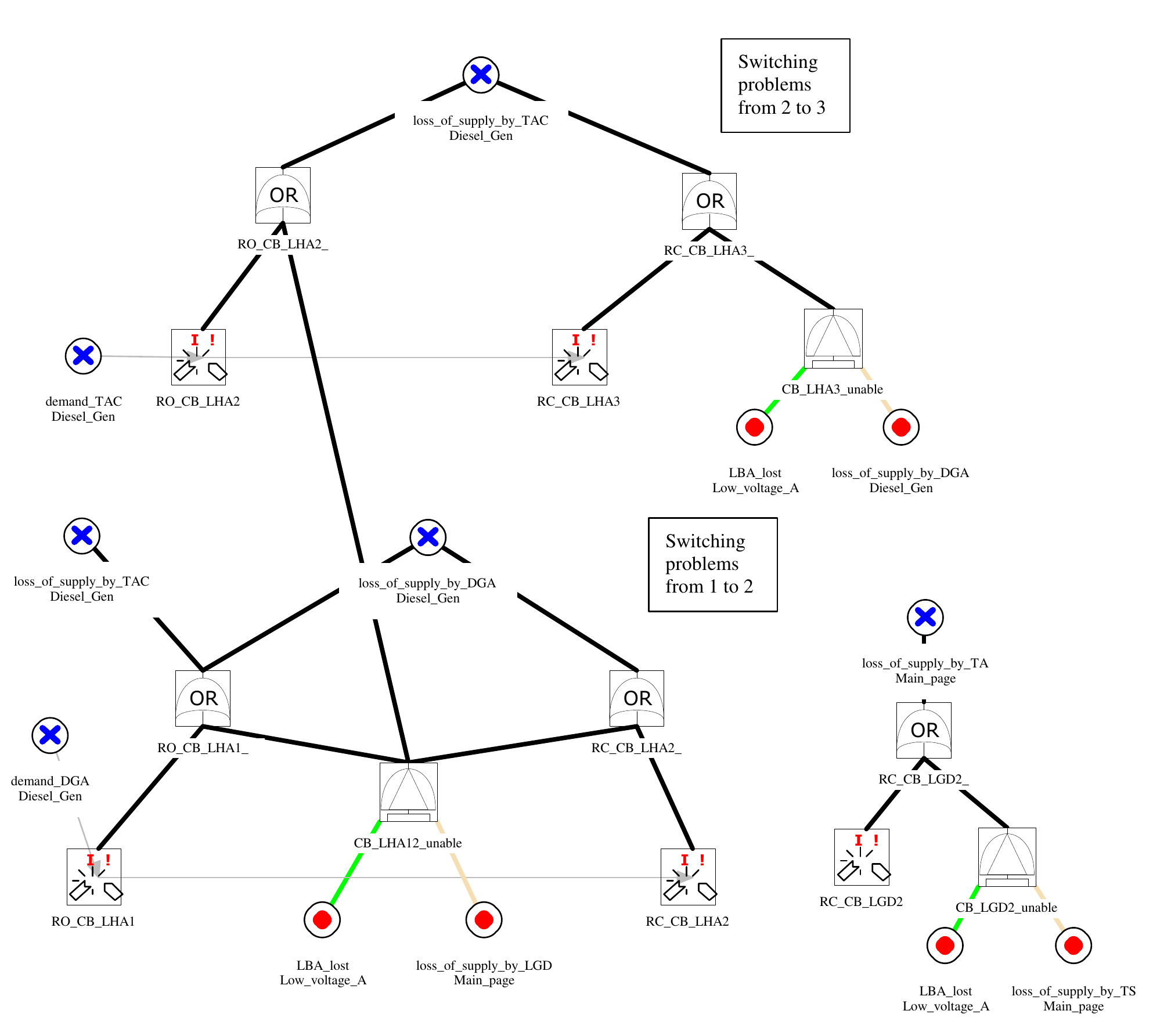}
\centering
\caption{Page Circuit\_breakers\_A.}
\end{figure}

\begin{figure}[h]
\includegraphics[width=0.9\textwidth]{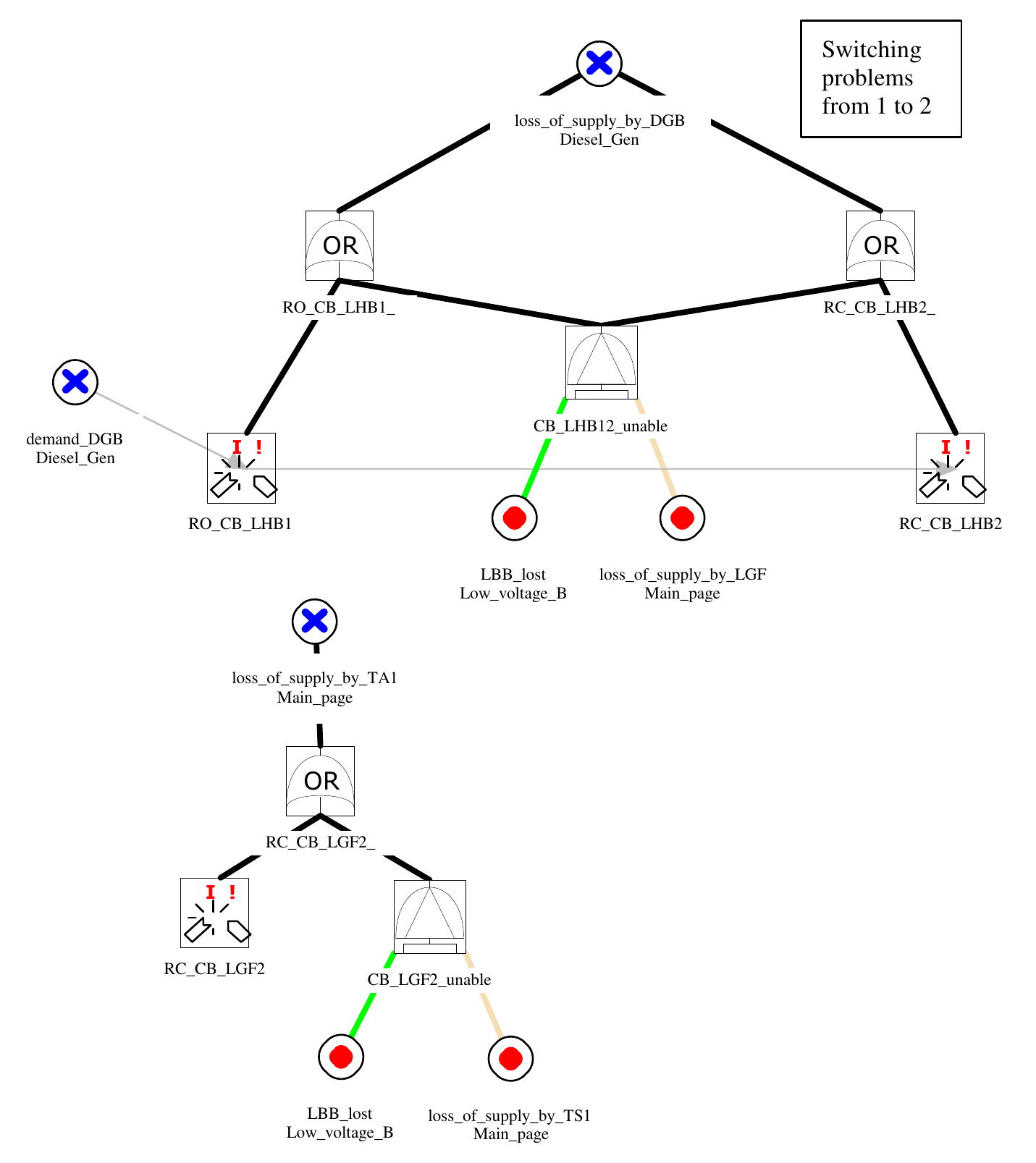}
\centering
\caption{Page Circuit\_breakers\_B.}
\end{figure}

\end{document}